\documentstyle[a4,12pt,epsf]{article}

\begin{document}

 \parindent 0pt \parskip \medskipamount

\newcommand{\elf}{ELFE}
\newcommand{\beq}{\begin{equation}}
\newcommand{\eeq}{\end{equation}}
\newcommand{\beqa}{\begin{eqnarray}}
\newcommand{\eeqa}{\end{eqnarray}}
\newcommand{\half}{\frac{1}{2}}
\newcommand{\gsim}{\buildrel > \over {_\sim}}
\newcommand{\lsim}{\buildrel < \over {_\sim}}
\newcommand{\ie}{{\it ie}}
\newcommand{\eg}{{\it eg}}
\newcommand{\cf}{{\it cf}}
\newcommand{\etal}{{\it et al.}}
\newcommand{\gev}{{\rm GeV}}
\newcommand{\jpsi}{J/\psi}
\newcommand{\order}[1]{${\cal O}(#1)$}
\newcommand{\morder}[1]{{\cal O}(#1)}
\newcommand{\eq}[1]{Eq.\ (\ref{#1})}
\newcommand{\ptr}{p_T}
\newcommand{\as}{\alpha_s}
\newcommand{\ket}[1]{\vert{#1}\rangle}
\newcommand{\bra}[1]{\langle{#1}\vert}
\newcommand{\cpair}{c\bar c}

\newcommand{\PL}[3]{Phys.\ Lett.\ {{\bf#1}} ({#2}) {#3}}
\newcommand{\NP}[3]{Nucl.\ Phys.\ {{\bf#1}} ({#2}) {#3}}
\newcommand{\PR}[3]{Phys.\ Rev.\ {{\bf#1}} ({#2}) {#3}}
\newcommand{\PRL}[3]{Phys.\ Rev.\ Lett.\ {{\bf#1}} ({#2}) {#3}}
\newcommand{\ZP}[3]{Z. Phys.\ {{\bf#1}} ({#2}) {#3}}
\newcommand{\PRe}[3]{Phys.\ Rep.\ {{\bf#1}} ({#2}) {#3}}

\begin{titlepage}
\begin{flushright}
        NORDITA--97/8 P\\
        hep-ph/9702385\\
       \today
\end{flushright}

\vskip 2.5cm

\centerline{\Large \bf Charmonium Production at ELFE Energies\footnote{Talk
given at the {\em Second ELFE Workshop}, Saint Malo, September 23-27, 1996.
Work supported in part by the EU/TMR contract ERB FMRX-CT96-0008.}}

\vskip 1.5cm

\centerline{\bf Paul Hoyer}
\centerline{\sl Nordita}
\centerline{\sl Blegdamsvej 17, DK--2100 Copenhagen \O, Denmark}

\vskip 2cm

\begin{abstract}
\noindent
I discuss issues related to charmonium production, in view of physics
possibilities at a 15 \ldots\ 30 GeV continuous beam electron facility. High
energy photo- and hadroproduction of heavy quarkonia presents several
challenges to QCD models concerning cross sections, polarization and nuclear
target dependence. Theoretical approaches based on color evaporation as well
as on color singlet and color octet mechanisms have met with both successes and
failures, indicating that charmonium production is a sensitive probe
of color dynamics. Experiments close to charm kinematic threshold will be
sensitive also to target substructure since only unusual, compact target
configurations contribute. In particular, subthreshold production on nuclei
should identify nuclear hot spots of high energy density. At low energies,
charmonium will form inside the target nucleus, allowing a determination of
$\cpair$ bound state interactions in nuclear matter.
\end{abstract}

\end{titlepage}

\newpage
\renewcommand{\thefootnote}{\arabic{footnote}}
\setcounter{footnote}{0}
\setcounter{page}{1}

\subsubsection*{1.\ \ \ Introduction}

The `Electron Laboratory for Europe' (\elf) project aims at a high intensity
$(I_e \simeq 30 \mu A)$ continuous electron beam in the 15 \ldots\ 30 GeV
energy range, to be scattered off fixed targets ranging from hydrogen to heavy
nuclei \cite{elfref}. Among the central physics motivations are detailed
studies of hadron and nuclear wave functions through hard exclusive and
(semi-)inclusive processes \cite{elfphys}. As it turns out, \elf\ will
operate in the region of charm $(\cpair)$ threshold, which in the case of
real photons is at $E_{\gamma}^{th} \simeq 8\ldots 12$ GeV for $\jpsi,\ldots,
D\bar D$ production on free nucleons. Charm(onium) production has proved to
be a very sensitive measure of reaction mechanisms, as evidenced by
order-of-magnitude discrepancies found between QCD models and data
\cite{schulrev,sansoni,mangano}. Furthermore, the suppression of charmonium
production in heavy ion collisions is widely discussed as a potential signal
for the formation of a quark-gluon plasma \cite{satz}. Here I would
like to discuss some of the puzzles of charmonium production and
how photo- and electroproduction close to threshold can give important new
clues to production mechanisms and to hadron and nuclear structure
\cite{frastri}.

\begin{figure}[htb]
\begin{center}
\leavevmode
{\epsfxsize=13.5truecm \epsfbox{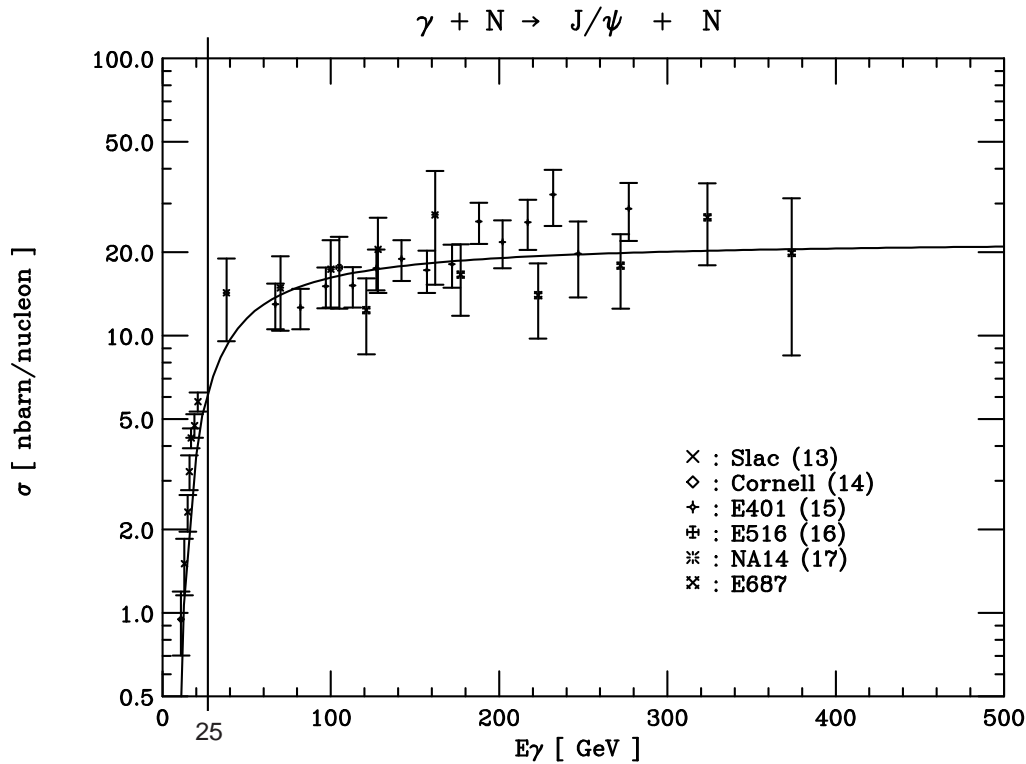}}
\end{center}
\caption[*]{Compilation of cross sections for the process $\gamma p\to \jpsi
p$ \cite{e687}. Experiments at ELFE will be in the range $E_\gamma\lsim 25$
GeV (vertical line). The curve shows the prediction of \eq{pgf} for a gluon
structure function $G(x)= 3(1-x)^5$.}
\label{fig1}
\end{figure}

Remarkably, the only charm photoproduction data that exists (Fig. 1) in the
\elf\ energy range are the $\jpsi$ measurements of SLAC \cite{slac} and
Cornell \cite{cornell} from 1975, which predate the discovery of open
charm. These early measurements of the small near-threshold cross section
$\sigma(\gamma N \to \jpsi N) \simeq 1$ nb were made possible by the
experimental cleanliness of the $\jpsi \to \mu\mu$ signal. With an \elf\
luminosity ${\cal L} \sim 10^{35} {\rm cm^{-2}s^{-1}}$ one expects a rate of
about 5 $\jpsi$ dimuon decays per second, allowing detailed measurements of
threshold and subthreshold effects. 

It should also be kept in mind that owing to the
essentially non-relativistic nature of charmonium, each charm quark
carries close to one half of the $\jpsi$ momentum. Even their relative angular
momentum is determined through the quantum numbers of the charmonium state.
Charmonium is thus a very valuable complement to open charm
channels such as $D\bar D$, which furthermore are difficult to
measure.

Theoretically, charmonium offers very interesting challenges. Most reliable
QCD tests have so far concerned hard inclusive scattering, implying a sum
over a large number of open channels. The standard QCD
factorization theorem \cite{fact} does not apply when the final
state is restricted by requiring the charm quarks to bind as charmonium.
The application of QCD to charmonium production is thus partly an art, as
evidenced by lively discussions of different approaches. It seems likely
that charmonium production will teach us something qualitatively new about
QCD effects in hard scattering -- exactly what is not yet clear (but
hopefully will be so by the time \elf\ turns on!).

\subsubsection*{2.\ \ \ $\jpsi$ Production at High Energies}

\medskip
{\sc 2.1\ \  Elastic $\jpsi$ Production}
\medskip

Early studies \cite{cem} assumed that the charmonium cross section is
proportional to the $\cpair$ one below open charm $(D\bar D)$ threshold, as
given by the inclusive photon-gluon fusion process $\gamma g \to \cpair$.
Thus
\beq
\sigma(\gamma N \to \jpsi+X) = f_{\jpsi} \int_{4m_c^2}^{4m_D^2} \frac{dM^2}{s}
G(M^2/s) \sigma_{\gamma g \to \cpair}\,(M^2)  \label{pgf}
\eeq
where $G(x)$ is the gluon structure function and the proportionality constant
$f_{\jpsi}$ is the fraction of the below-threshold $\cpair$ pairs that
form $\jpsi$'s. In this `Color Evaporation Model' (CEM) the
color exchanges which transform the color octet $\cpair$ pair into a
color singlet $\jpsi$ are assumed to occur over long time and distance
scales, and are described by the non-perturbative factor $f_{\jpsi}$ in
\eq{pgf}. For the model to have predictability it is important that this factor
be `universal', \ie, independent of the reaction kinematics (beam
energy and charmonium momentum), and hopefully also of the nature of the
projectile and target. It should be emphasized, however, that the universality
of $f_{\jpsi}$ is a hypothesis which has not been demonstrated in QCD.

Assuming a constant $f_{\jpsi}$ and a `standard' gluon structure function
$xG(x)=3(1-x)^5$, \eq{pgf} (with $X=N$) gives a good fit (solid line in Fig. 1)
to $\jpsi$ elastic photoproduction from threshold to $E_\gamma \lsim
300$ GeV \cite{e687}. It is not very clear what this means, however. Close
to threshold the single gluon exchange picture is expected to break down
(\cf\ section 3.1). At high energy, color evaporation is expected to apply to
{\em inelastic} processes, since the neutralization of color will lead to
additional hadrons being produced.

\begin{figure}[htb]
\begin{center}
\leavevmode
{\epsfxsize=13.5truecm \epsfbox{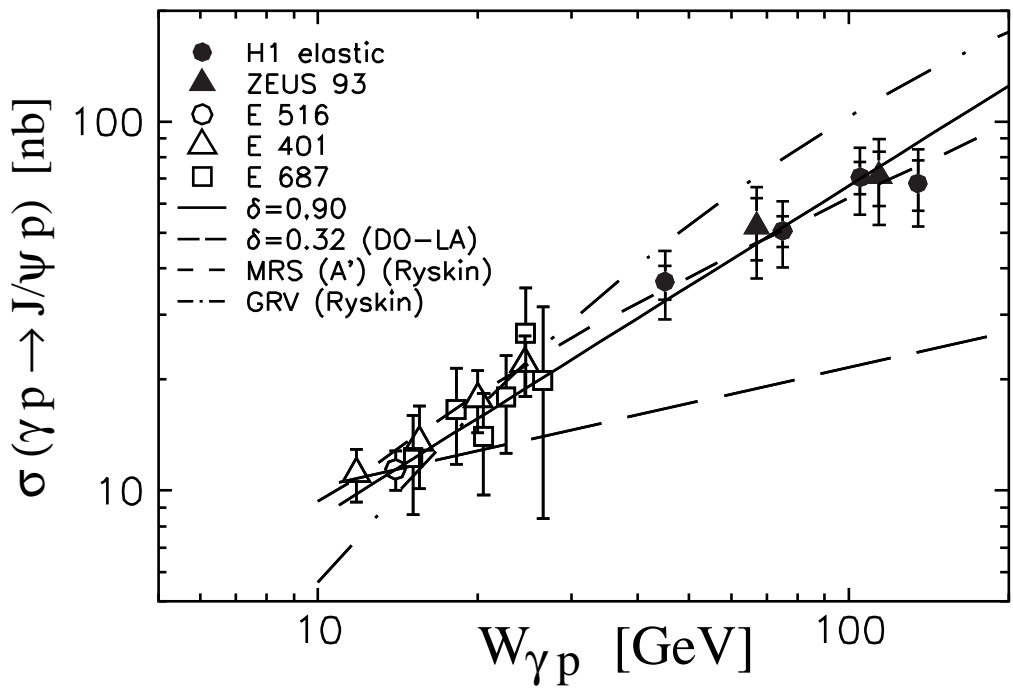}}
\end{center}
\caption[*]{Compilation of high energy data on the process $\gamma p\to \jpsi
p$ \cite{h1}, with curves of the form $W_{\gamma p}^\delta$ as indicated.
The curves marked `MRS' and `GRV' are the results of QCD calculations with
two-gluon exchange \cite{elqcd,rrml}, for different gluon structure functions.}
\label{fig2}
\end{figure}

A consistent QCD description of high energy elastic $\jpsi$ photoproduction
involves two gluon (color singlet) exchange between the charm quark pair and
the target \cite{elqcd}. In this approach the color dynamics of the charm
quark pair is treated perturbatively, \ie, the quarks are created as a compact
color singlet state which couples directly to the $\jpsi$ through the wave
function at the origin. This is justified by a factorization between the hard
and soft physics in this process \cite{elfact}. The (imaginary part of the) two
gluon coupling to the target resembles the gluon structure function of deep
inelastic scattering, although the momenta of the two gluons are not exactly
equal \cite{radyushkin}. The elastic $\jpsi$ cross section may thus be
approximately proportional to the {\em square} of the gluon structure
function. The high energy data (Fig. 2) on $\gamma p \to \jpsi p$ from HERA
\cite{zeus,h1} in fact shows a considerable rise of the elastic cross section
with energy, which (within the considerable error bars) is consistent with the
increase of $G(x)$ for $x \simeq 4m_c^2/s \to 0$ \cite{rrml}.

\bigskip
{\sc 2.2\ \  Color Evaporation Approach to Inelastic $\jpsi$ Production}
\medskip

The difficulties of perturbative QCD models in describing the data on
inelastic charmonium production (\cf\ sections 2.3 and 2.4 below) has
rekindled interest in the color evaporation model
\cite{gavai,schuler,amundson,schulvogt}. It has been shown that the dependence
of both charmonium and bottomonium production on the projectile energy and on
the energy fraction $x_F$ of the produced state are in good agreement with
that predicted through \eq{pgf} for heavy quarks below threshold. The fraction
of the below-threshold heavy quark cross section which ends up in quarkonium
depends on the QCD parametrization (quark mass, structure functions and
factorization scale) but seems to be quite small, typically (8 \ldots\ 10)\%
for charm, growing to (17 \ldots\ 32)\% for bottom \cite{schulvogt}. The
parameter $f_{\jpsi}$ of \eq{pgf} takes similar values in $pp$ and
$\pi p$ reactions (0.025 and 0.034, respectively \cite{gavai}). In
photoproduction the large diffractive (elastic) peak needs to be excluded, \eg,
by a cut on the $\jpsi$ momentum, after which values in the range $f_{\jpsi}=
0.005\ldots 0.025$ were found \cite{schulvogt}.

The generally good agreement of the color evaporation model with data is very
significant. It shows that the essential structure of the inclusive charmonium
cross section is given by that of heavy quark production at leading twist.
According to the spirit of color evaporation, the heavy quarks will after
their production undergo a long time-scale process of evolution to the
quarkonium bound state, during which the relative distance between the quarks
grows and non-perturbative gluons change the overall color of the
quark pair. The normalization of the production cross section, \ie, the
non-perturbative parameter $f_{\jpsi}$ in \eq{pgf}, is thus not necessarily
related to the wave function at the origin of the charmonium bound state.

\bigskip
{\sc 2.3\ \ The Color Singlet Model (CSM)}
\medskip

The `Color Singlet Model' (CSM) \cite{csm} describes charmonium
production fully in terms of PQCD. The $\cpair$ is created with proper quantum
numbers to have an overlap with the charmonium state, measured by the
non-relativistic wave function at the origin. In particular, the pair has to
be a singlet of color. For inelastic $\jpsi$ photoproduction the lowest order
subprocess is $\gamma g \to \cpair g$, where the final gluon radiation
ensures that the charmonium is produced with an energy fraction (in the
target rest frame) $z=E_{\jpsi}/E_\gamma<1$. For production at large $p_\perp
\gg m_c$ higher order 'fragmentation diagrams' actually give the leading
contribution \cite{frag}.

\begin{figure}[htb]
\begin{center}
\leavevmode
{\epsfxsize=13.5truecm \epsfbox{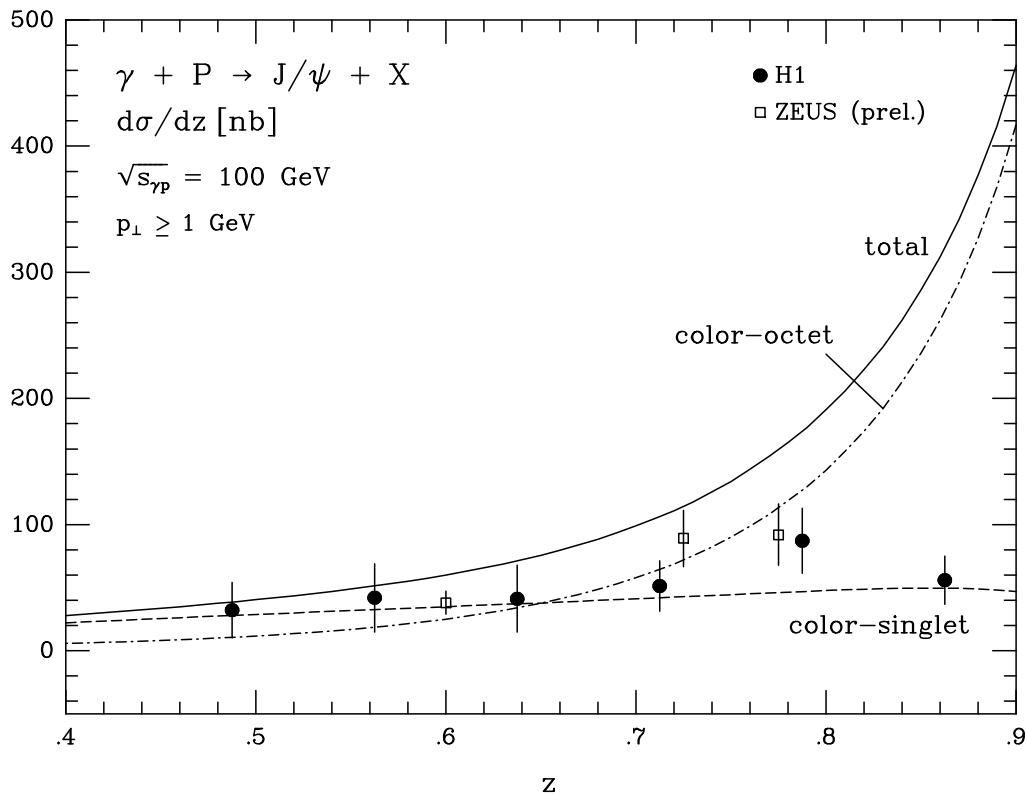}}
\end{center}
\caption[*]{The cross section for inelastic $\jpsi$ photoproduction $\gamma p
\to \jpsi+X$ for $p_\perp(\jpsi) \ge 1$ GeV as a function of the $\jpsi$ energy
fraction (in the proton rest frame) $z=E_{\jpsi}/E_\gamma$ \cite{cakr}.
Predictions based on the color singlet and octet mechanisms are compared to
data from HERA.}
\label{fig3}
\end{figure}

The CSM contributions to $\jpsi$ photoproduction have been calculated to
next-to-leading order in QCD \cite{kramer}, with a result that is in good
agreement with the data (Fig. 3). For $\psi'$ production the CSM predicts
that the $\psi'/\psi$ cross section ratio should be proportional to the square
of the wave function at the origin,
\beq
\frac{\sigma(\psi')}{\sigma_{dir}(\jpsi)}=\frac{\Gamma(\psi'\to
\mu\mu)}{M_{\psi'}^3}\frac{M_{\jpsi}^3}{\Gamma(\jpsi\to \mu\mu)}
 \simeq .24\pm .03  \label{dirratio}
\eeq
Here $\sigma_{dir}(\jpsi)$ excludes contributions to the $\jpsi$ from
`indirect' channels such as $B$, $\chi_c$ and $\psi'$ decays, and the power of
mass is motivated by dimensional arguments. The photoproduction data
\cite{na14,nmc} gives for the ratio that is uncorrected for radiative decays,
\beq
\frac{\sigma(\gamma N \to \psi'+X)}{\sigma(\gamma N \to \jpsi+X)}=0.20\pm 0.05
\pm 0.07  \label{fullratio}
\eeq
The upper limit on the $\chi_{c1}+\chi_{c2}$ photoproduction cross section is
about 40\% of the $\jpsi$ cross section \cite{na14}. Taking into account the
\order{20\%} branching ratio for their radiative decays into $\jpsi$ only a
small fraction of the photoproduced $\jpsi$'s are due to the indirect channels,
and the ratios of Eqs. (\ref{dirratio}) and (\ref{fullratio}) should be
compatible, as indeed they are. It should be noted, however, that the
experimental ratio (\ref{fullratio}) primarily reflects diffractive (elastic)
$\jpsi$ and $\psi'$ production, which dominates in photoproduction.

In hadroproduction, where inelastic channels dominate the cross section,
data on the ratio (\ref{dirratio}) is also in good agreement with the
color singlet model \cite{vhbt}. This is
true also for the Tevatron data on charmonium production at large $p_\perp$
\cite{sansoni}. Even bottomonium production is quite consistent with the
analog of
\eq{dirratio}, within factor two uncertainties due to the so far unmeasured
contributions from radiative decays of the P states \cite{gavai}.

The above comparisons suggest that the `nonperturbative' proportionality
factors $f$ in the color evaporation model (\cf\ \eq{pgf}) actually reflect
perturbative physics, \ie, the wave function at the origin as assumed in
the color singlet model.

In spite of its successful predictions in photoproduction and of the {\em
ratio} of $\psi'$ to $\jpsi$ hadroproduction, the CSM nevertheless
fails badly, by factors up to 30 \ldots\ 50, for the {\em
absolute hadroproduction} cross sections of the $\jpsi$, the $\psi'$ and
the $\chi_{c1}$ states \cite{schulrev,sansoni,mangano,vhbt}. The discrepancies
are large both in fixed target total cross section data and in
large $p_\perp$ production at the Tevatron. The fixed target data moreover
shows that the $\jpsi$ and $\psi'$ are produced nearly unpolarized
\cite{fixpol}, contrary to the CSM which predicts a fairly large transverse
polarization \cite{vhbt}.

The fact that the CSM {\em underestimates} charmonium hadroproduction (and
predicts the polarization incorrectly), suggests that there are other
important production mechanisms, beyond the CSM. The nature of those
processes is not yet established. A simple mnemonic, which appears to be
consistent with the observed systematics, is that the CSM works whenever no
extra gluon emission is required only to satisfy the quantum number
constraints. Thus, for inelastic $\jpsi$ photoproduction the lowest order
process $\gamma g \to \cpair g$ of the CSM has only the number of gluons which
is required by momentum transfer (and the prediction is successful). In the
(incorrect) CSM prediction for hadroproduction  gluon
emission in the subprocess $gg\to \cpair g$ is needed only due to the negative
charge conjugation of the $\jpsi$ (or due to Yang's theorem in the case of
$\chi_{c1}$ production). Again, for $\chi_{c2}$ the lowest order process $gg
\to \chi_{c2}$ is allowed in the CSM, and the prediction is compatible with
the data (within the considerable PQCD uncertainties)
\cite{mangano,vhbt,cgmp}. A polarization measurement of hadroproduced
$\chi_{c2}$'s would be a valuable check of the CSM \cite{vhbt,chipol}.

Photoproduction of $\chi_c$ is an interesting test case
\cite{schulvogt,cakr}. The available data \cite{na14} suggests that the
$\chi_{c2}/\jpsi$ ratio is lower in inelastic photoproduction than in
hadroproduction. This qualitatively agrees with the CSM, in which $P$-wave
photoproduction $(\gamma g\to \chi_{c2} gg)$ is of higher order than $S$-wave
production $(\gamma g\to \jpsi g)$, while the reverse is true for
hadroproduction. With no regard to quantum numbers (as in the color
evaporation model) the basic subprocess would be the same ($\gamma g
\to \cpair$) and the $\chi_{c}/\jpsi$ ratio would be expected to be similar
in photo- and hadroproduction.

It has been suggested that the gluons required to satisfy quantum number
constraints of the $\cpair$ pair in the CSM could come from additional (higher
twist) exchanges with the projectile or target \cite{vhbt,htwist}. Although
normally suppressed, these contributions might be important since they do not
involve energy loss through gluon emission.

\bigskip
{\sc 2.4\ \ The Color Octet Model (COM)}
\medskip

A possible solution to some of the above puzzles has been
suggested based on an analysis of nonrelativistic QCD (NRQCD) \cite{nrqcd},
and commonly referred to as the `Color Octet Model' (COM) \cite{cgmp,com}. In
cases where, due to quantum number constraints, extra gluon emission is
required in the CSM the production may be dominated instead by higher order
terms in the relativistic ($v/c$) expansion of the quarkonium bound state.
For $P$-wave states the inclusion of relativistic corrections is in fact
necessary to cancel infrared divergencies of the perturbative expansion even
at lowest order.

The $\cpair$ can then be produced in a color octet state, which has
an overlap with a higher $\ket{\cpair g}$ Fock state of charmonium, with the
emission of a soft gluon. Such contributions appear in a systematic NRQCD
expansion and thus must exist. Whether they are big enough to account for the
large discrepancies of the CSM in charmonium production depends on the
magnitude of certain non-perturbative matrix elements of NRQCD. I refer to
recent reviews \cite{mangano,annrev} and references therein to the
extensive literature on this subject.

A number of discrepancies between the color octet model and observations
suggest that it will at best provide only a partial explanation of quarkonium
production.
\begin{itemize}

\item[{\em (i)}] Inelastic photoproduction of $\jpsi$ is overestimated by the
COM \cite{cakr,photoprod} as seen in Fig. 3. A best estimate of the discrepancy
is actually even larger than shown, since the effects of soft gluon
radiation were neglected in fitting the octet matrix elements from the
Tevatron data \cite{softg}. The photoproduction cross section can be {\em
decreased} in the COM only by adding a contribution which is coherent with
the production amplitude.

\item[{\em (ii)}] The COM does not explain the $p_\perp$-integrated (fixed
target) charmonium hadroproduction data, in particular not the polarization of
the $\jpsi$ and $\psi'$ and the $\chi_{c1}/\chi_{c2}$ ratio \cite{tv,bz,gs}.
It has been claimed (but is by no means obvious) that the (higher twist?)
corrections are bigger in the fixed target data than in the large $p_\perp$
cross section measured at the Tevatron, which is often taken as a benchmark
for COM fits. The systematics of the anomalies is actually very similar in
the two processes. The color singlet model fails by a comparable
factor in both cases \cite{sansoni}, while the (leading twist) color
evaporation model successfully explains the relative production rates measured
in the fixed target and Tevatron experiments \cite{amundson,schulvogt}.

\item[{\em (iii)}] The $\Upsilon(3S)$ cross section exceeds the CSM
predictions by an order of magnitude \cite{sansoni,cdfups}. Since the
relativistic corrections are much smaller for bottomonium than for charmonium,
this is hard to accomodate in the COM \cite{bz}. It has been suggested that
the excess could be due to radiative decay from a hitherto undiscovered $3P$
state. As in the case of the `$\psi'$ anomaly', an experimental measurement
of direct $\Upsilon$ production should settle this question. The ratios of
$\Upsilon(nS)$ cross sections are quite compatible with expectations based on
the wave function at the origin (\cf\ \eq{dirratio}), with only moderate
contributions from radiative decays of $P$ states \cite{gavai}. 
\end{itemize}

\bigskip
{\sc 2.5\ \  Nuclear Target $A$-Dependence}
\medskip

Additional clues to quarkonium production dynamics is offered by data
on the nuclear target $A$-dependence \cite{hoyrev}. In the standard
parametrization
\beq
\sigma(A) \propto A^{\alpha}  \label{adep}
\eeq
one expects $\alpha \simeq 1$ for hard incoherent scattering, which is
additive on all nucleons in the target nucleus. This behavior is verified
with good precision for the Drell-Yan process of large-mass lepton pair
production \cite{dy} as well as for open charm ($D$ meson) production at low
$x_F$ \cite{dprod}. However, for $\jpsi$ and $\psi'$ hadroproduction $\alpha
\simeq 0.92 \pm .01$ for $.1 \lsim x_F \lsim .3$ \cite{cha}. This suppression may be
interpreted as a rescattering of the charm quark pair in the nucleus, with an
effective cross section of 7 mb \cite{satz} for conversion to open charm
production. Such rescattering will affect the quantum numbers of the
$\cpair$ pair, and should thus be considered in color singlet and octet
approaches. For the color evaporation model the target dependence shows that
the proportionality factor $f_{\jpsi}$ in \eq{pgf} is not universal for all
processes.

The nuclear suppression of $\jpsi$ and $\psi'$ production increases with
$x_F$, with $\alpha(x_F=.6) \simeq .8$ \cite{cha}. This effect, which breaks
leading twist factorization \cite{hvs}, may be due to intrinsic charm
\cite{ic,icpsi} and involve the scattering of low momentum valence quarks
\cite{bhmt}. Ascribing the effect to parton energy loss in the nucleus
requires the $\langle p_\perp \rangle$ in the rescattering to be unexpectedly
large \cite{bh,jr}. The dynamics of charmonium production at large $x_F$ is
analogous to the large $p_\perp$ Tevatron data due to the `trigger bias'
effect: In both cases the charmonium carries a large fraction of the momentum
of the fragmenting particle.

In inelastic (virtual) photoproduction a nuclear {\em enhancement} of $\jpsi$
production is observed,  $\alpha =1.05\pm 0.03$ for $x_F < 0.85$ and
$p_\perp^2 > 0.4\ \gev^2$ \cite{nmc,e691}. Contrary to hadroproduction, the
momentum distribution of $\jpsi$ photoproduction peaks at large $x_F$. Hence
an explanation in terms of energy loss is conceivable \cite{hkz}.

In the region of the coherent peak for $\jpsi$ photoproduction on nuclei at
very low $p_\perp$ there is an even stronger nuclear enhancement. E691
\cite{e691} finds $\alpha_{coh} = 1.40 \pm 0.06 \pm 0.04$, while NMC
\cite{nmc} gives $\alpha_{coh}=1.19\pm 0.02$. If the $\cpair$ pairs are
 compact enough not to suffer secondary interactions in the nucleus, one
expects $\sigma_{coh}(A, p_\perp) \propto A^2 \exp(-cA^{2/3}p_\perp^2)$ ($c$
being a constant). Hence $\alpha_{coh}=4/3$ for the $p_\perp$-integrated cross
section, in rough agreement with the data.

\medskip

As should be clear from the above, quarkonium production offers
interesting challenges, which are not fully met by any one proposed mechanism.
It seems likely that we are learning something about color dynamics that
cannot be accessed within the standard, fully inclusive formalism of PQCD.
Color exchanges to the $\cpair$ evidently take place in ways not adequately
described by the CSM. The importance of the NRQCD contributions (which surely
are present at some level) remains to be clarified, as does the assumption by
the color evaporation approach that charmonium production constitutes
a universal fraction of the $\cpair$ cross section below open charm threshold.

\subsubsection*{3.\ \ \ Production Near Kinematic Threshold}

As noted in the Introduction, almost all experimental information on
charmonium production is at relatively high energy. While we may hope that
at least some of the puzzles discussed in the preceding section will be solved
in the near future, an understanding of production near threshold will
have to wait for a dedicated machine like ELFE. In the following I shall
discuss some generic features of (sub-)threshold charmonium production
related to the composite nature of the beam and/or target\footnote{
Calculations of higher order perturbative, leading twist effects in heavy quark
production near threshold may be found in Ref. \cite{thr}.} \cite{frastri}. It
is likely that charm production close to threshold will teach us new
physics, over and beyond what is now being learnt at higher energies.

\bigskip
{\sc 3.1\ \ Higher Twist Effects}
\medskip

At high energy the dominant contribution to hard processes comes from
`leading twist' diagrams, characterized by only one parton from each
colliding particle participating in the large momentum transfer $(Q)$
subprocess. Since the time scale of the hard collision is $1/Q$, only
partons within this transverse distance can affect the process. The
likelihood that two partons are found so close to each other is typically
proportional to the transvers area $1/Q^2$, which thus gives the suppression
of higher twist, multiparton contributions.

Close to the kinematic boundary the higher twist effects are enhanced,
however. Thus for $\gamma p \to \cpair p$ very near threshold, all the
partons of the proton have to transfer their energy to the charm quarks
within their creation time $1/m_c$, and must thus be within this transverse
distance from the $\cpair$ and from each other. The longitudinal momentum
transfer at threshold (in the proton rest frame) is $\simeq m_c$. Hence only
compact proton Fock states, with a radius equal to the compton wavelength of
the heavy quark, can contribute to charm production at threshold.

\begin{figure}[htb]
\begin{center}
\leavevmode
{\epsfxsize=13.5truecm \epsfbox{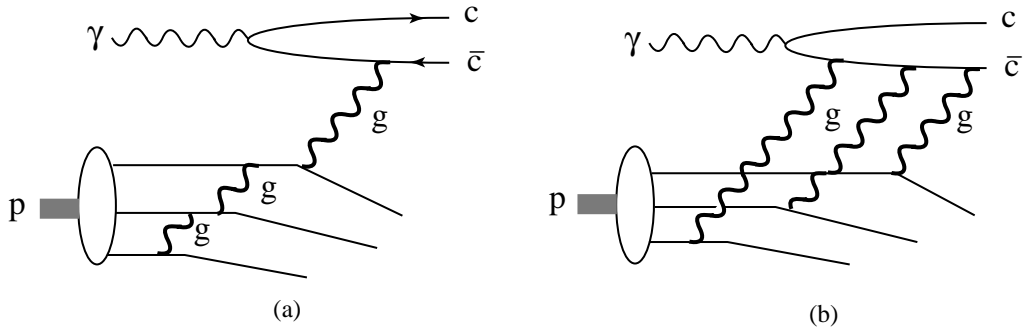}}
\end{center}
\caption[*]{Two mechanisms for transferring most of the proton momentum to the
charm quark pair in $\gamma p \to \cpair + X$ near kinematic threshold. The
leading twist contribution (a) dominates at high energies, but becomes
comparable to the higher twist contribution  (b) close to threshold.}
\label{fig4}
\end{figure}

The behavior of the effective proton radius in charm photoproduction near
threshold can be surmised from the following argument. As indicated in Fig.
4a, one mechanism for charm production is that most of the proton momentum is
first transferred to one (valence) quark, followed by a hard subprocess
$\gamma q \to \cpair q$. If the photon energy is $E_\gamma = \zeta
E_\gamma^{th}$, where $E_\gamma^{th}$ is the energy at kinematic threshold
($\zeta \gsim 1$), the valence quark must carry a fraction $x=1/\zeta$ of
the proton's (light-cone) momentum. The lifetime of such a Fock state
(in a light-cone or infinite momentum frame) is $\tau \simeq 1/\Delta E$,
where 
\beq
\Delta E= \frac{1}{2p}\left[m_p^2-\sum_i \frac{p_{i\perp}^2+m_i^2}{x_i}
\right] \simeq -\frac{\Lambda_{QCD}^2}{2p(1-x)}  \label{dele}
\eeq
For $x=1/\zeta$ close to unity such a short-lived fluctuation can be created
(as indicated in Fig. 4a) through momentum transfers from valence proton
states (where the momentum is divided evenly) having commensurate
lifetimes $\tau$, \ie, with
\beq
r_\perp^2 \simeq \frac{1}{p_\perp^2} \simeq
\frac{\zeta-1}{\Lambda_{QCD}^2} \label{psize}
\eeq
This effective proton size thus decreases towards threshold $(\zeta \to 1)$,
reaching $r_\perp \simeq 1/m_c$ at threshold, $\zeta-1 \simeq
\Lambda_{QCD}^2/m_c^2$.

As the lifetimes of the contributing proton Fock states
approach the time scale of the $\cpair$ creation process, the time ordering of
the gluon exchanges implied by Fig. 4a ceases to dominate
higher twist contributions such as that of Fig. 4b \cite{bhmt}, which are
related to intrinsic charm \cite{ic}. There are in fact reasons to expect that
the latter diagrams give a dominant contribution to charmonium production near
threshold. First, there are many more such diagrams. Second, they allow the
final state proton to have a small transverse momentum (the gluons need
$p_\perp \simeq m_c$ to couple effectively to the $\cpair$ pair, yet the
overall transfer can still be small in Fig. 4b). Third, with several gluons
coupling to the charm quark pair its quantum numbers can match those of a
given charmonium state without extra gluon emission.

The above discussion is generic, and does not indicate how close to threshold
the new effects actually manifest themselves. While more quantitative model
calculations certainly are called for, this question can only be
settled by experiment. It will be desirable to measure both the cross section
and polarization for several charmonium states, as well as for
open charm. At present, there are only tantalizing indications for novel
phenomena at charm threshold, namely:
\begin{itemize}
\item {\em Fast $\cpair$ pairs in the nucleon.}
The distribution of charm quarks in the nucleon, as measured by deep inelastic
lepton scattering, appears \cite{emc} to be anomalously large at high $x$,
indicating a higher twist intrinsic charm component \cite{ic}. An analogous
effect is suggested by the high $x_F$ values observed in $\pi N\to \jpsi+
\jpsi +X$ \cite{twopsi}. A proton Fock state containing charm quarks with a
large fraction of the momentum will enhance charm production close to
threshold. 

\item {\em $\jpsi$ polarization in $\pi^- p\to \jpsi+X$ for $x_F\to 1$.}
Only compact projectile $(\pi)$ Fock states contribute in the limit
where the $\jpsi$ carries almost all of the projectile momentum \cite{berbro}.
It may then be expected that the helicity of the $\jpsi$ equals the helicity of
the projectile, \ie, the $\jpsi$ should be longitudinally polarized. This
effect is observed both in the above reaction \cite{fixpol} and in $\pi N \to
\mu^+\mu^- + X$ \cite{e615}.

\item {\em Polarization in $pp \to pp$ large angle scattering.} There is a
sudden change in the $A_{NN}$ polarization parameter close to charm threshold
for 90$^\circ$ scattering \cite{ann}. It has been suggested that this is due to
an intermediate state containing a $\cpair$ pair, which has low angular
momentum due to the small relative momenta of its constituents
\cite{broter}. This idea could be tested at ELFE by investigating correlations
between polarization effects in large angle compton scattering, $\gamma p \to
\gamma p$, and charm production ($\gamma p \to \cpair p$) near
threshold.

\item {\em Change in color transparency at charm threshold.} Intermediate
states with a charm quark pair could also give rise to the sudden decrease
in color transparency observed in $pA \to pp(A-1)$ close to charm threshold
\cite{coltra}. Due to the low momentum of the constituents they expand to a
large transverse size within the nucleus, thus destroying transparency
\cite{broter}. Again, $\gamma A$ reactions could provide important tests at
ELFE.
\end{itemize}

\bigskip
{\sc 3.2\ \ Subthreshold production}
\medskip

The high luminosities at ELFE will allow detailed studies of subthreshold
production of charm(onium). It is well established that antiprotons and kaons
are produced on nuclear targets at substantially lower energies than is
kinematically possible on free nucleons \cite{subth}. Thus the minimal
projectile energy required for the process $pp\to \bar p+X$ on free protons
at rest is 6.6 GeV, while the kinematic limit for $pA\to \bar p+X$ on a heavy
nucleus at rest is only $3m_N\simeq 2.8$ GeV. Antiproton production has been
observed in $p+\, ^{63}$Cu collisions down to $E_{lab}^p\simeq 3$ GeV, very
close to kinematic threshold. Scattering on a single nucleon in the nucleus
would at this energy require a fermi momentum of \order{800} MeV. While the
$pA$ data can be fit assuming such high Fermi momenta, this assumption leads
to an underestimate of subthreshold production in $AA$ collisions by about
three orders of magnitude
\cite{fermom}.

\begin{figure}[htb]
\begin{center}
\leavevmode
{\epsfxsize=13.5truecm \epsfbox{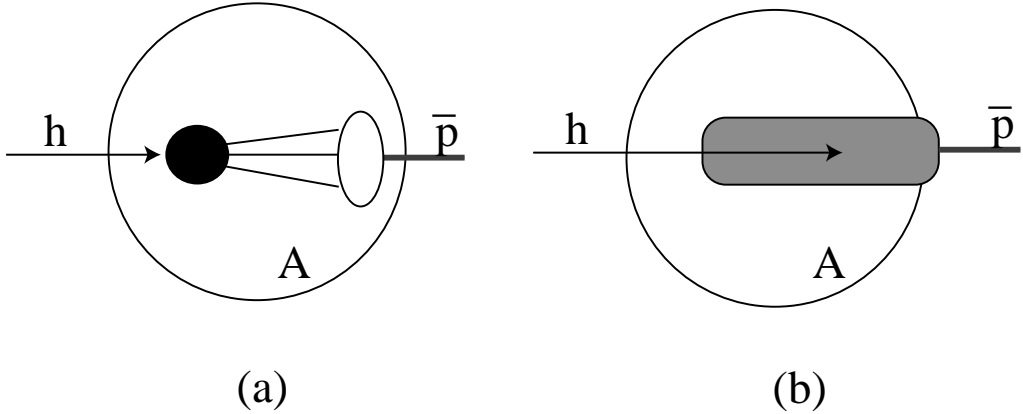}}
\end{center}
\caption[*]{Two conceptual mechanisms for subthreshold $\bar p$ production in
$hA$ collisions. In (a) the production occurs locally off a hot spot (black
circle) of high energy density in the nucleus. In (b) the light quarks gain
momentum over an extended nuclear region (grey).}
\label{fig5}
\end{figure}

There are at least two qualitatively different scenarios for the observed
subthreshold production of antiprotons. Either (Fig. 5a) the projectile
strikes a local `hot spot' with a high energy density in the nucleus. The
effective mass of the scatterer is high, lowering the
kinematic threshold. Alternatively (Fig. 5b) the momentum required
to create the antiproton is not transferred locally, but picked up in an
extended longitudinal region: the nucleus forms a `femtoaccelerator'.
Establishing either scenario would teach us something qualitatively new about
rare, highly excited modes of the nucleus.

Real and virtual photoproduction of charm below threshold would be of crucial
help in distinguishing the correct reaction mechanism, for several reasons. 

\begin{itemize}
\item The photon is pointlike, and is thus a clean probe
of target substructure. In particular, effects due to the shrinking effective
size of a hadron probe near threshold (\cf\ discussion above) are eliminated.

\item The $\cpair$ pair is created locally, within a proper time $\tau \simeq
1/m_c$. The extended acceleration scenario of Fig. 5b is thus not effective for
charm production. If significant subthreshold charm production occurs (beyond
what can be ascribed to standard fermi motion) this selects the hot spot
scenario of Fig. 5a.

\item Subthreshold production can be studied as a function of the virtuality
$Q^2$ of the photon. Little $Q^2$ dependence is expected for $Q^2 \lsim
m_c^2$, due to the local nature of charm production. Nuclear hot spots
smaller than $1/m_c$ would be selected at higher values of $Q^2$.
\end{itemize}

\bigskip
{\sc 3.3\ \ Interactions of $\cpair$ Pairs in Nuclei}
\medskip

Close to threshold for the process $\gamma p \to \jpsi p$ on stationary
protons the energy of the $\jpsi$ is $E_{\jpsi}^{lab} \simeq 7$ GeV. This
corresponds to a moderate lorentz $\gamma$-factor $E_{\jpsi}/M_{\jpsi}
\simeq 2.3$. Hence a significant expansion of the $\cpair$ pair occurs
inside large nuclei, and effects of charmonium bound states in nuclei may be
explored.

Compared to the propagation of light quarks in nuclei, charm has the
advantage that one can readily distinguish hidden (charmonium) from open
$(D\bar D)$ charm production. Thus the dependence of the
$\sigma(\jpsi)/\sigma(D)$ ratio on the target size $A$ and on projectile
energy indicates the amount of rescattering in the nucleus. The
presently available data on the $A$-dependence of charmonium production is
at much higher energies (\cf\ section 2.5), and thus measures the nuclear
interactions of a compact $\cpair$ pair rather than of full-sized charmonium.
Further information about the significance of the radius of the charmonium
state can be obtained by comparing $\psi'$ to $\jpsi$ production on various
nuclei. In high energy $hA$ and $\gamma A$ scattering both states have very
similar $A$-dependence \cite{cha}.

Information about the propagation of charmonium in nuclei is very important
also for relativistic heavy ion collisions, where charmonium production may
be a signal for quark-gluon plasma formation \cite{satz}. Precise information
from ELFE would allow a more reliable determination of the background signal
from charmonium propagation in ordinary nuclear matter.

Even though the $\cpair$ pair is created with rather high momentum even at
threshold, it may be possible to observe reactions where the pair is captured
by the target nucleus, forming `nuclear-bound quarkonium' \cite{bts}. This
process should be enhanced in subthreshold reactions. There is no Pauli
blocking for charm quarks in nuclei, and it has been estimated there is a large
attractive van der waals potential binding the pair to the nucleus \cite{lms}.
The discovery of such qualitatively new bound states of matter would be a scoop
for any accelerator.

\subsubsection*{4.\ \ \ Summary}

Charmonium production provides a valuable window to color dynamics. The
experimental signal from the $\jpsi$ dilepton decay mode is very clean,
allowing the measurement of a small signal in the presence of much more
abundant background processes. The charmonium production process is hard due
to the large mass of the charm quark. The binding energy is low, however,
making charmonium a sensitive probe a color exchange processes. 

Charmonium photoproduction was measured close to threshold soon after the
discovery of the $\jpsi$ \cite{slac,cornell}. The more recent experiments have
all been done well above threshold, and have revealed a number of
puzzles, including `anomalous' production rates, polarization and nuclear
target $A$-dependence.

The Color Evaporation Model \cite{cem} assumes that charmonium cross
sections are universal fractions of $\cpair$ production. It successfully
describes the $E_{CM}$, $x_F$ and $p_\perp$ dependence of $\jpsi$ production,
and in particular correctly predicts the relative magnitude of the Tevatron
high $p_\perp$ data and the low $p_\perp$ data of fixed target
experiments. The universality seems to be broken, however, in $\chi_c$
photoproduction and for scattering on nuclear targets.

The Color Singlet Model \cite{csm} successfully predicts (within the
typically rather large uncertainties of charm production) the absolute
normalization for $\jpsi$ photoproduction and $\chi_{c2}$ hadroproduction.
The model fails by more than an order of magnitude in $\jpsi$ and $\chi_{c1}$
hadroproduction, however.

The Color Octet Model \cite{com} is based on a systematic NRQCD expansion in
the relative velocity $v/c$ of the heavy quarks. The higher order terms in
$v/c$ involve new non-perturbative parameters, which can be determined from a
fit to the Tevatron $\jpsi$ data at high $p_\perp$. A fully consistent picture
has yet to emerge. In particular, this approach overestimates $\jpsi$
photoproduction, underestimates $\Upsilon(3S)$ production and fails to
describe the fixed target data on $\jpsi$ polarization.

ELFE will provide precise data on charmonium production close to threshold.
New effects arise since only compact Fock states of the target can
contribute to near-threshold production. Analogous effects have been
observed at high energies through a change in the polarizaton state of the
$\jpsi$ for $x_F \to 1$.

In scattering on nuclei, subthreshold charmonium production will be sensitive
to nuclear `hot spots', compact subclusters of high energy density. Such
clusters could give rise to the observed high rates of subthreshold $\bar p$
and $K$ production \cite{subth}, but light hadrons need not necessarily be
created locally in the nucleus.

The relatively low momenta of the $\cpair$ pairs produced close to threshold
will allow studies of charmonium (rather than of compact $\cpair$) interactions
in the nucleus. Such interactions may explain the anomalous effects observed
in the polarization and color transparency of large angle $pp$ elastic
scattering. It may even be possible to study $\cpair$ pairs at rest in nuclear
matter, where they could form a new form of matter, nuclear bound quarkonium.

{\bf Acknowledgements.} I am grateful to the organizers of this meeting for
their invitation to discuss physics in a stimulating atmosphere and a
delightful setting. I also wish to thank S. Brodsky and M. V\"anttinen, in
particular, for longstanding collaborations on the issues presented
above, as well as B. Kopeliovich and M. Strikman for several useful
discussions.

\end{document}